\documentclass[seceq,preprint]{ptptex}

\usepackage[dvips]{graphicx}

  \newcommand{\nn}{\nonumber}
\def\Av{\mbox{\boldmath $A$}}

\def\Cv{\mbox{\boldmath $C$}}

\def\Kv{\mbox{\boldmath $K$}}

\def\qv{\mbox{\boldmath $q$}}
\def\kv{\mbox{\boldmath $k$}}

\preprintnumber[4cm]{
KUNS-2171\\
YNU-HEPTh-08-102\\
KEK-CP-219}

\title{
Heavy Quark Effects in the Virtual Photon Structure Functions
}

\author{
 Yoshio   \textsc{Kitadono}$^{1}$,
 \footnote{e-mail: kitadono@scphys.kyoto-u.ac.jp}
 Ken      \textsc{Sasaki}$^{2}$,
 \footnote{e-mail: sasaki@phys.ynu.ac.jp}
 Takahiro \textsc{Ueda}$^{3}$,
 \footnote{e-mail: uedat@post.kek.jp}\\
 and 
 Tsuneo  \textsc{Uematsu}$^{4}$
 \footnote{e-mail: uematsu@scphys.kyoto-u.ac.jp}
}

\inst{

${}^{1}$ Dept. of Physics, Faculty of Science, Hiroshima University,\\ 
Higashi Hiroshima 739-8526, Japan.\\

${}^{2}$ Dept. of Physics, Faculty of Engineering, 
Yokohama National University,\\ 
Yokohama 240-8501, Japan.\\

${}^{3}$ High Energy Accelerator Research Organization (KEK), \\
1-1 Oho, Tsukuba, Ibaraki 305-0801, Japan.\\

${}^{4}$ Dept. of Physics, Graduate School of Science, Kyoto University,\\ 
Yoshida, Kyoto 606-8501, Japan.\\
}

\abst{
We investigate the heavy quark mass effects in the virtual photon 
structure functions $F_{2}^{\gamma}(x, Q^2, P^2)$ and 
$F_{L}^{\gamma}(x, Q^2, P^2)$ in the framework of the 
mass-independent renormalization group equation (RGE). 
We study a formalism in which the heavy quark mass effects are treated
based on parton picture as well as on the operator product expansion (OPE), 
and perform the numerical evaluation of 
$F_{\rm eff}^{\gamma}(x, Q^2, P^2)$ to the next-leading order (NLO)
in QCD.
}
\begin{document}

\maketitle


\section{Introduction}
The  Large Hadron Collider (LHC)\cite{LHC} has started its operation and it is 
anticipated that the signals for the new physics beyond the Standard Model
 (SM) will be discovered.
Once these signals are observed, more precise measurements will need to be 
carried out at the future $e^+e^-$  collider, so-called the International 
Linear Collider (ILC) \cite{ILC}.
In such cases, it is still important for us to have detailed knowledge of 
the SM predictions at high energies based on QCD.

\begin{figure}
 \centerline{\includegraphics[width=7.5cm,height=5cm]
{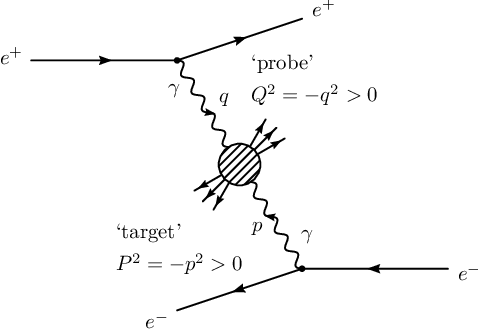}}
\caption{Deep-inelastic scattering on a virtual photon in the
$e^+e^-$ collider experiments}
\label{fig1}
\end{figure}

It is well known that, in high energy $e^+ e^-$ collision experiments, the 
cross section of the two-photon processes $e^+ e^- \rightarrow e^+ e^- + 
{\rm hadrons}$ dominates over that of the one-photon annihilation processes 
$e^+ e^- \rightarrow \gamma^*\rightarrow {\rm hadrons}$. 
The two-photon 
processes provide a good testing ground for studying the predictions of QCD 
at high energies. Here we consider the two-photon processes in the double-tag 
events where both of the outgoing $e^+$ and $e^-$ are detected 
(see Fig.~\ref{fig1}). 
In particular, we investigate the kinematical region  in which one of the 
photons with momentum $q$ is far off-shell (large $Q^2\equiv -q^2$) while 
the other with momentum $p$ is close to the mass-shell (small $P^2=-p^2$), 
can be viewed as a deep-inelastic scattering where the target is a photon 
rather than a nucleon~\cite{WalshBKT}. 
In this deep-inelastic scattering off photon targets, we can study the photon 
structure functions, which are the analogs of the nucleon structure functions. 

The unpolarized (spin-averaged) photon structure functions $F_2^\gamma(x, Q^2)$
as well as
$F_L^\gamma(x, Q^2)$ of the real photon ($P^2\!=\!0$) were first studied in
the parton-model (PM) \cite{PM1_real}
and then investigated in perturbative QCD (pQCD). The 
leading order (LO) QCD contributions to $F_2^\gamma(x, Q^2)$ and 
$F_L^\gamma(x, Q^2)$ were obtained by Witten \cite{real_photon_LO}. 
The next-to-leading order 
(NLO) corrections to $F_2^\gamma(x, Q^2)$ were calculated by Bardeen and 
Buras \cite{real_photon_NLO}. These results were obtained in the framework 
based on the operator product 
expansion (OPE) \cite{OPE} and the renormalization group equation 
(RGE)\cite{RGE}. 
The same results were rederived by the QCD improved PM \cite{QCDPM}.

The structure function $F_2^\gamma(x,Q^2,P^2)$ for the case of a virtual 
photon target ($P^2\ne 0$) was investigated in the  LO and in the NLO 
by pQCD \cite{virtual_photon_LO,virtual_photon_NLO}. 
Also the virtual photon structure function $F_L^\gamma(x,Q^2,P^2)$ was studied 
in the LO \cite{virtual_photon_NLO}. 
In fact, these structure functions were analyzed
in the kinematical region, $\Lambda^2 \ll P^2 \ll Q^2$, where $\Lambda$ is the 
QCD scale parameter. The advantage of studying a virtual photon target in this 
kinematical region is that we can calculate the whole structure function, its 
shape and magnitude, by the perturbative method. This is contrasted with the 
case of the real photon target where in the NLO there exist nonperturbative 
pieces.
Recently the QCD analysis was made for $F_2^\gamma(x,Q^2,P^2)$ up to the 
next-to-next-to-leading order (NNLO) and for 
$F_L^\gamma(x,Q^2,P^2)$ up to the NLO \cite{virtual_photon_NNLO}.  
And more recently the target effects on these structure functions were 
studied \cite{virtual_photon_TMC} and compared with the existing experimental 
data\cite{PLUTO,L3}.
In these calculations all the relevant quarks were assumed to be massless.

In this paper we examine the heavy quark mass effects on the photon structure 
functions $F_2^\gamma(x,Q^2,P^2)$ and $F_L^\gamma(x,Q^2,P^2)$. 
Indeed, the heavy quark mass effects for the two-photon processes, especially 
in the deep-inelastic kinematical region, have been studied by many
authors \cite{Gluck-Reya,positivity,Schienbein,CJK,New5}. 
But they were not treated within the framework of the OPE and the RGE. 
Our analysis here is performed in the framework of the QCD improved PM 
powered by  the parton evolution equations and based on the mass-independent 
renormalization group approach in which the RGE parameters, i.e., $\beta$ and 
$\gamma$ functions, are the same as those of the massless quark case. 
We consider the system which consists of $n_f-1$ massless quarks and one 
heavy quark together with  gluons and photons. Then, the heavy quark mass 
effects are included in the RGE inputs; the coefficient functions and the 
operator matrix elements. 
In the case of the nucleon target, the heavy quark mass effects were studied 
by a method  based on the OPE in Ref.~\citen{Buza-etal}, where the heavy quark 
was  treated such that it was radiatively generated and  absent in the
intrinsic quark components of the nucleon. This picture does not hold
for the case of virtual photon target, since the heavy quark is also generated from the virtual 
photon target with light quarks at high energies. We should consider both the heavy and light quarks 
equally  as the partonic components inside the virtual photon.

In the next section, we derive the evolution equations for the parton
distribution functions in the case where $n_f-1$ light quarks and one heavy 
quark are present.
In section 3, we calculate the heavy quark effects in the virtual 
photon structure functions, $F_{2}^{\gamma}(x,Q^2,P^2)$, 
$F_{L}^{\gamma}(x,Q^2,P^2)$ and $F_{\rm eff}^{\gamma}(x,Q^2,P^2)$,
and compare  our
theoretical predictions with the existing experimental data.
The final section is devoted to the conclusions.
We discuss the diagonalization of the anomalous dimension in Appendix A, 
and the parton-model derivation of the master formula in Appendix B. 


\section{The evolution equations in the presence of the heavy quark effects}
\label{derive1}
In this section we consider the evolution equations for the case in which
$n_f-1$ massless quarks and one heavy quark exist.
The extension to the system with many heavy quarks is 
straightforward, since we can repeat this treatment recursively.

\subsection{Heavy quark effects and operator mixing}
We discuss the heavy quark effects in the RGE mixing and derive the 
master formula for the moments with quark mass effects in this subsection.
Although the OPE is the useful formalism, we can get more physically intuitive
picture in the PM approach, so we discuss the heavy quark effects in the
evolution equations in the parton model.
Now we consider the evolution equations for $n_{f}-1$ massless quark
parton distribution functions (PDFs) $q^{i}(x,Q^2,P^2)$ 
($i=1,\cdots,n_f-1$) and one heavy quark PDF $q^{H}(x,Q^2,P^2)$
together with the gluon PDF $G^\gamma(x,Q^2,P^2)$
and photon PDF $\Gamma^{\gamma}(x,Q^2,P^2)$,
Experimentally, this situation corresponds to system of
u,d,s (massless quarks) + c (heavy quark) for kinematical region of 
PLUTO data\cite{PLUTO} and u,d,s,c (massless quarks) + b (heavy) 
for that of L3\cite{L3}.
We write down the DGLAP equations for $q^{i}$, $q^{H}$, $G^{\gamma}$,
$\Gamma^\gamma$; 
\begin{eqnarray}
&& \frac{dq^{i}(x,Q^2,P^2)}{d \ln Q^2} \nn\\
&&= \int_{1}^{x} \frac{dy}{y} 
     \left[   \sum_{j=1}^{n_{f}-1} \tilde{P}_{ij}
              \left(\frac{x}{y}, Q^2\right) q^{j}(y, Q^2, P^2)
	    + \tilde{P}_{iH}
              \left(\frac{x}{y}, Q^2\right) q^{H}(y, Q^2, P^2)
\right. \nn\\
&&{} \left. \hspace{2.0cm}
	    + \tilde{P}_{qG}
              \left(\frac{x}{y}, Q^2\right) G^{\gamma}(y,Q^2,P^2)
	    + \tilde{P}_{i\gamma}
              \left(\frac{x}{y}, Q^2\right) \Gamma^{\gamma}(y,Q^2,P^2)
     \right], \label{DGLAP_q}
\\
&& \frac{dq^{H}(x,Q^2,P^2)}{d \ln Q^2} \nn\\
&&= \int_{1}^{x} \frac{dy}{y} 
     \left[   \sum_{j=1}^{n_{f}-1} \tilde{P}_{Hj}
              \left(\frac{x}{y}, Q^2\right) q^{j}(y, Q^2, P^2)
	    + \tilde{P}_{HH}
              \left(\frac{x}{y}, Q^2\right) q^{H}(y, Q^2, P^2)
\right. \nn\\
&&{} \left. \hspace{2.0cm}
	    + \tilde{P}_{HG}
	      \left(\frac{x}{y}, Q^2\right) G^{\gamma}(y, Q^2, P^2)
	    + \tilde{P}_{H\gamma}
              \left(\frac{x}{y}, Q^2\right) \Gamma^{\gamma}(y,Q^2,P^2)
     \right], \label{DGLAP_H}
\\
&& \frac{d G^{\gamma}(x,Q^2,P^2)}{d \ln Q^2}\nn\\
&&= \int_{1}^{x} \frac{dy}{y} 
     \left[   \sum_{j=1}^{n_{f}-1} \tilde{P}_{Gq}
              \left(\frac{x}{y}, Q^2\right) q^{j}(y, Q^2, P^2)
	    + \tilde{P}_{GH}
              \left(\frac{x}{y}, Q^2\right) q^{H}(y, Q^2, P^2)
\right. \nn\\
&&{} \left. \hspace{2.0cm}
	    + \tilde{P}_{GG}
	      \left(\frac{x}{y}, Q^2\right) G^{\gamma}(y, Q^2, P^2)
	    + \tilde{P}_{G\gamma}
              \left(\frac{x}{y}, Q^2\right) \Gamma^{\gamma}(y,Q^2,P^2)
     \right]. \label{DGLAP_G}
\end{eqnarray}
where $\tilde{P}_{ij} = \delta_{ij}\tilde{P}_{qq} 
+ \frac{1}{n_f}\tilde{P}_{qq}^{S}$ is the splitting functions of 
$j$-parton into $i$-parton, and the first term represents the
process that $j$-quark splits into $i$-quark without through
gluon, and the second term stands for the splitting through gluon
and $\widetilde{P}_{qq}$ and $\widetilde{P}_{qq}^S$ are both
independent of quark flavors, $i$ and $j$. $\widetilde{P}_{qq}^S$
is relevant for the flavor-singlet part and starts in the order
of $\alpha_s^2$.

We now define the singlet combination, $q^{\gamma}_L$, and the
non-singlet part of $i$-th quark $q^{i}_{NS}$ as well as the
non-singlet combination $q^{\gamma}_{NS}$ for the
light-flavors as,
\begin{eqnarray}
 q^{\gamma}_{L} &\equiv& \sum_{i=1}^{n_{f}-1} q^{i}, \quad
 q^{i}_{NS} \equiv q^{i} - \frac{1}{n_{f}-1} q^{\gamma}_{L},\quad
 q^{\gamma}_{NS} \equiv \sum_{i=1}^{n_f-1} e_{i}^2 q^{i}_{NS}.
\label{singlet-NS}
\end{eqnarray}
Note that $\sum_{i=1}^{n_f-1}q^{i}_{NS}=0$.
We should remember that the photon PDF in the virtual photon 
$\Gamma^{\gamma}$ does not evolve within the order $ \alpha_{QED}$,
so we set $\Gamma^{\gamma}(y,Q^2,P^2)=\delta(1-y)$.
We can rewrite the Eqs.(\ref{DGLAP_q}), (\ref{DGLAP_H}) and
 (\ref{DGLAP_G}) 
in terms of $q^{\gamma}_{L}, q^{\gamma}_{NS}$. This can be done by the 
following steps.
At first, we sum up Eqs.(\ref{DGLAP_q}),(\ref{DGLAP_H}) and (\ref{DGLAP_G}) 
from $i=1$ to $i= n_f-1$ and we get the equation 
for the parton distribution defined as
a row vector $\qv^{\gamma}(x,Q^2,P^2)=(q^{\gamma}_{L}, q^{H}, G^{\gamma},
q^{\gamma}_{NS})$,
\begin{eqnarray}
  \frac{d}{d \ln Q^2} \qv^{\gamma}(x,Q^2,P^2)
 &=& \int_{1}^{x} \frac{dy}{y} 
     \left[   \qv^{\gamma}(y,Q^2,P^2) 
              \hat{P}\left(\frac{x}{y}, Q^2\right) 
     \right]
 + \kv(x,Q^2,P^2),
\end{eqnarray}
where the splitting functions are given by a matrix
\begin{eqnarray}
\hat{P} &\equiv&
 \begin{pmatrix}
   P_{qq}^{S} &  P_{LH} &  P_{LG} & 0           \\
   P_{HL}     &  P_{HH} &  P_{HG} & 0           \\
   P_{GL}     &  P_{GH} &  P_{GG} & 0           \\
   0          &  0      &  0      & P_{qq}^{NS} 
 \end{pmatrix}~.
\end{eqnarray}
with
\begin{eqnarray}
 P_{qq}^{S} &=& \tilde{P}_{qq} + \frac{n_f - 1}{n_f} \tilde{P}_{qq}^{S},
  \hspace{0.3cm}
 P_{LH}      =  \frac{n_f - 1}{n_f} \tilde{P}_{qq}^{S},
  \hspace{0.3cm}
 P_{LG}      =  (n_f - 1)\tilde{P}_{qG},\nn \\
 P_{HL}^{S} &=& \frac{1}{n_f} \tilde{P}_{qq}^{S},
  \hspace{0.3cm}
 P_{HH}      =  \tilde{P}_{qq} + \frac{1}{n_f} \tilde{P}_{qq}^{S},
  \hspace{0.3cm}
 P_{HG}      =  \tilde{P}_{HG}, \nn\\
 P_{GL}^{S} &=& \tilde{P}_{Gq}^{S},
  \hspace{0.3cm}
 P_{GH}      =  \tilde{P}_{GH},
  \hspace{0.3cm}
 P_{GG}      =  \tilde{P}_{GG}, \hspace{0.5cm}
 P_{qq}^{NS} =  \tilde{P}_{qq}~.  
\end{eqnarray}
and the inhomogeneous term, 
$\kv \equiv  \left( k_{L}, k_{H}, k_{G}, k_{NS} \right)$ 
describing the parton-photon mixing is given by
\begin{eqnarray}
k_{L} &=& \sum_{i=1}^{n_f-1}\tilde{P}_{i\gamma}, 
 \hspace{0.3cm}
k_{H}  = \tilde{P}_{H\gamma},
 \hspace{0.3cm}
k_{G}  = \tilde{P}_{G\gamma},
 \hspace{0.3cm}
k_{NS} = \sum_{i=1}^{n_f-1}e_{i}^2
         \left(   \tilde{P}_{i\gamma} 
	        - \frac{1}{n_f-1}\sum_{j=1}^{n_f-1} \tilde{P}_{j\gamma} 
	 \right).\nn\\
\end{eqnarray}
Note that the moments of the splitting functions $\tilde{P}_{ij}$ are 
related to the anomalous dimensions of operators $\gamma_n(g)$ and the 
coefficient
function $C_n^i$ satisfies the following mass-independent RGE:
\begin{eqnarray}
\left[\mu\frac{\partial}{\partial\mu}+\beta(g)\frac{\partial}{\partial g}
+\gamma_m(g){m}\frac{\partial}{\partial m}-{\gamma}_n(g,\alpha)\right]_{ij}
C_n^j\left(\frac{Q^2}{\mu^2},\frac{m^2}{\mu^2},{\bar g}(\mu^2),\alpha\right)
=0~,
\end{eqnarray}
where $\gamma_m(g)$ is the anomalous dimension for the mass operator.
The solution to this equation is given by
\begin{eqnarray}
C_n^i\left(\frac{Q^2}{\mu^2},\frac{m^2}{\mu^2},{\bar g}(\mu^2),\alpha\right)
=\left\{T\exp
\left[ \int^{\bar{g}(\mu^2)}_{\bar{g}(Q^2)}dg 
	             \frac{\gamma_{n}(g,\alpha)}{\beta(g)}\right]
\right\}_{ij}
C_n^j\left(1,\frac{{\bar m}^2}{Q^2},{\bar g}(Q^2),\alpha\right)~,
\nonumber\\
\end{eqnarray}
where the anomalous dimension $\gamma_n(g,\alpha)$ is a $5 \times 5$ matrix
and is given by
\begin{eqnarray}
\gamma_{n}(g,\alpha) \equiv 
 \begin{pmatrix}
  \hat{\gamma}_n    & 0\\
  \Kv_n&0
 \end{pmatrix}, \quad
 {\hat{\gamma}_n}\equiv
\begin{pmatrix}
   \gamma_{LL}^{n} &  \gamma_{HL}^{n} &  \gamma_{GL}^{n} & 0                \\
   \gamma_{LH}^{n} &  \gamma_{HH}^{n} &  \gamma_{GH}^{n} & 0                \\
   \gamma_{LG}^{n} &  \gamma_{HG}^{n} &  \gamma_{GG}^{n} & 0                \\
     0             &    0         &    0         & \gamma_{NS}^{n}  \\
\end{pmatrix}~,\label{anomalous-dim}
\end{eqnarray}
where 
\begin{eqnarray}
\Kv_n=(K^n_L,K^n_H,K^n_G,K^n_{NS})~.
\end{eqnarray}
which describes the mixing between hadronic operators and 
the photon operator.

\subsection{Master formula for the moment}
We can summarize our master formula for the $n$-th moment of the virtual
photon structure functions in the case where a heavy quark exists as
\begin{eqnarray}
 M^{\gamma}_{2}(n, Q^2, P^2)
 &=& \sum_{i,j=\psi,H,G,NS,\gamma} 
 A^{i}_{n}\left(1, \frac{\bar{m}^2(P^2)}{P^2}, \bar{g}(P^2) \right) 
   \left\{{T}\exp \left[ \int^{\bar{g}(P^2)}_{\bar{g}(Q^2)}dg 
	             \frac{\gamma_{n}(g,\alpha)}{\beta(g)}  
	      \right]\right\}_{ij} \nn\\
  &{}& \hspace{1.0cm} 
      \times C^{j}_{2,n}\left(1, \frac{\bar{m}^2(Q^2)}{Q^2},
			 \bar{g} (Q^2)\right)~,\label{moment-mass}
\end{eqnarray}
where $\psi$ ($NS$) is the flavor singlet (non-singlet) quark operators
for the $n_f-1$ massless quarks. 
$H$ stands for the heavy quark and
$\bar{m}(Q^2)$ is the running mass evaluated at $Q^2$. 
$A_{n}^{i}$ is the operator matrix element renormalized at 
$\mu^2=P^2$, while $C_{2,n}^{i}$ is the coefficient function
renormalized at $\mu^2=Q^2$.
Since $A_{n}^{i}, C_{2,n}^{i}$ are the 
solutions of the renormalization group equation, 
they depend on the running masses $\bar{m}(Q^2)$ and $\bar{m}(P^2)$.
We can decompose the moments Eq.~(\ref{moment-mass}) into that for
the massless case and  the additional term 
$\Delta M_{2}^{\gamma}(n, Q^2, P^2)$ due to the mass effects:
\begin{eqnarray}
 M_{2}^{\gamma}(n, Q^2, P^2)
 &=& M_{2}^{\gamma}(n, Q^2, P^2){\Bigr\vert}_{\rm massless} + \Delta M_{2}^{\gamma}(n, Q^2, P^2)~.
\end{eqnarray}
They are obtained up to NLO by the diagonalization of the anomalous 
dimension matrix,
which will be discussed in Appendix \ref{derive1}.

The master formula for the $n$-th moment to NLO is given by
\begin{eqnarray}
&{}& M_{2}^{\gamma}(n, Q^2, P^2) 
  = \int_{0}^{1}dx x^{n-2} F_{2}^{\gamma}(x,Q^2,P^2)\nn\\ 
 &=& \frac{\alpha}{4\pi}\frac{1}{2\beta_{0}}
     \left[   \frac{4\pi}{\alpha_s(Q^2)} \sum_{i} \mathcal{L}_{i}^{n}
              \left[ 1 - \left( \frac{\alpha_s(Q^2)}
			        {\alpha_s(P^2)} \right)^{d_{i}^n + 1} 
	      \right]
            + \sum_{i} \mathcal{A}_{i}^{n}
              \left[ 1 - \left( \frac{\alpha_s(Q^2)}
			        {\alpha_s(P^2)} \right)^{d_{i}^n } 
	      \right]
\right. \nn\\
&{}& \left. \hspace{2.4cm}
            + \sum_{i} \mathcal{B}_{i}^{n}
              \left[ 1 - \left( \frac{\alpha_s(Q^2)}
			        {\alpha_s(P^2)} \right)^{d_{i}^n+1 } 
	      \right]
\hspace{0.2cm} 
	    + \mathcal{C}^{n} 
\hspace{0.2cm} 
     \right] + \mathcal{O}(\alpha_s)~,\label{master-org}
\end{eqnarray}
where $\alpha$ ($\alpha_s$) is the QED (QCD) coupling constant.
The summation index $i$ runs over $\pm,NS$ corresponding to the
three eigenvalues $\lambda_\pm^n$ and $\lambda_{NS}^n$ of the one-loop
anomalous dimension matrix in the massless case. While for the 
present case with a heavy flavor, 
$i$ runs over $\psi,\pm,NS$ for the four eigenvalues of
the one-loop anomalous dimension matrix $\hat{\gamma}_n^{(0)}$ which 
turns out to be $\lambda_\psi^n$, $\lambda_\pm^n$, $\lambda_{NS}^n$, where 
we have $\lambda_\psi^n=\lambda_{NS}^n$ (See Appendix \ref{derive1}).
For the notation of the renormalization group parameters we refer to
Ref.~\citen{virtual_photon_NNLO}. The LO coefficients 
$\mathcal{L}^{n}_{i}$, the NLO coefficients
$\mathcal{A}_{i}^{n}, \mathcal{B}_{i}^{n},\mathcal{C}^{n}$ are,
\begin{eqnarray}
\mathcal{L}^{n}_{i} 
 &=& \Kv_{n}^{(0)} P_{i}^{n} \Cv_{2,n}^{(0)}
     \frac{1}{1+d_{i}^{n}}, \nn\\
\mathcal{A}^{n}_{i} 
 &=& - \Kv_{n}^{(0)} \sum_{j} 
       \frac{ P_{j}^{n} \hat{\gamma}_{n}^{(1)} P_{i}^{n} }
            {\lambda_{j}^{n} - \lambda_{i}^{n} + 2\beta_{0}}
	\Cv_{2,n}^{(0)} \frac{1}{ d_{i}^{n} } 
     - \Kv_{n}^{(0)} P_{i}^{n} \Cv_{2,n}^{(0)}
       \frac{\beta_1}{\beta_0} \frac{1 - d_{i}^{n}}{d_{i}^{n}} \nn\\
&{}& + \Kv_{n}^{(1)}  P_{i}^{n} \Cv_{2,n}^{(0)}
       \frac{1}{ d_{i}^{n} } 
     - 2 \beta_0 \tilde{\Av}_n^{(1)} P_{i}^{n}
         \Cv_{2,n}^{(0)}, \nn\\
\mathcal{B}^{n}_{i} 
 &=&   \Kv_{n}^{(0)} \sum_{j} 
       \frac{ P_{i}^{n} \hat{\gamma}_{n}^{(1)} P_{j}^{n} }
            {\lambda_{i}^{n} - \lambda_{j}^{n} + 2\beta_{0}}
	\Cv_{2,n}^{(0)} \frac{1}{ 1+d_{i}^{n}} 
     + \Kv_{n}^{(0)} P_{i}^{n} \Cv_{2,n}^{(1)}
       \frac{1}{1 + d_{i}^{n}} \nn\\
&{}& - \Kv_{n}^{(0)}  P_{i}^{n} \Cv_{2,n}^{(0)}
       \frac{\beta_1}{\beta_0}
       \frac{ d_{i}^{n} }{ 1 + d_{i}^{n} }, \nn\\
\mathcal{C}^{n} 
 &=& 2 \beta_0 \left(   C_{2,n}^{\gamma (1)} 
		      + \tilde{\Av}_n^{(1)} \cdot 
		        \Cv_{2,n}^{(0)} 
	       \right).
\end{eqnarray}  
where $\Kv_n^{(0)}$ ($\Kv_n^{(1)}$) is the 1-loop (2-loop) photon-parton 
mixing anomalous dimension, $P^n_i$'s are projection operators,
$\Cv_{2,n}^{(0)}$ ($\Cv_{2,n}^{(1)}$) is the tree-level (1-loop) coefficient
function and $\beta_0$ ($\beta_1$) is the 1-loop (2-loop) beta function.
$d^n_i=\lambda^n_i/2\beta_0$.  $\Av_n^{(1)}=(\alpha/4\pi)\tilde{\Av}_n^{(1)}$ 
is the
1-loop operator matrix element and $C_{2,n}^{\gamma(1)}$ is the coefficient
function of the photon operator. (See the Ref.~\citen{virtual_photon_NNLO} 
for details). 
We can reorganize the summation over $i=\psi,\pm,NS$ for the case with a 
heavy flavor, into that for $i=\pm,NS$ since the eigenvalue $\lambda_\psi^n$
gives rise to the same exponents in Eq.~(\ref{master-org}), as $\lambda_\psi^n
=\lambda_{NS}^n$. 
The additional terms arising from the variation of
OME and the coefficient functions due to the heavy quark effects are,
\begin{eqnarray}
&{}& \Delta M_{2}^{\gamma}(n, Q^2, P^2, m^2) 
 = \int_{0}^{1}dx x^{n-2} \Delta F_{2}^{\gamma}(x,Q^2,P^2, m^2)\nn\\ 
 &=& \frac{\alpha}{4\pi}\frac{1}{2\beta_{0}}
     \left[ \hspace{0.2cm}   
              \sum_{i=\pm,NS} \Delta \mathcal{A}_{i}^{n}
              \left[ 1 - \left( \frac{\alpha_s(Q^2)}
			        {\alpha_s(P^2)} \right)^{d_{i}^n } 
	      \right]
\right. \nn\\
&{}& \left. \hspace{1.2cm}
            + \sum_{i=\pm,NS} \Delta \mathcal{B}_{i}^{n}
              \left[ 1 - \left( \frac{\alpha_s(Q^2)}
			        {\alpha_s(P^2)} \right)^{d_{i}^n+1 } 
	      \right]
	    + \Delta \mathcal{C}^{n} 
\hspace{0.2cm} 
     \right] + \mathcal{O}(\alpha_s)~, \label{master}
\end{eqnarray}
where $\Delta \mathcal{A}_{i}^{n}, \Delta \mathcal{B}_{i}^{n}, 
\Delta \mathcal{C}^{n}$ are the deviations from the massless case
due to the heavy quark effects. Note that the heavy quark effects do not 
change the LO coefficients $\mathcal{L}_{i}^{n}$ (See Appendix \ref{derive1}). 
This can be derived by an
alternative method (See Appendix \ref{derive2}).
 The explicit expressions of $\Delta \mathcal{A}_{i}^{n}, 
\Delta \mathcal{B}_{i}^{n}, \Delta \mathcal{C}^{n}$ are
\begin{eqnarray}
\Delta \mathcal{A}_{NS}^{n} 
 &=& \frac{1}{n_f} e_{H}^2(-12 \beta_0) 
     \Delta \tilde{A}_{nG}^{\psi}
     ( e_{H}^2 - \langle e^2 \rangle_{n_f} ), \nn\\
\Delta \mathcal{A}_{\pm}^{n} 
 &=& \frac{1}{n_f} e_{H}^2(-12 \beta_0) \langle e^2 \rangle_{n_f}
     \Delta \tilde{A}_{nG}^{\psi}
     \frac{\gamma^{0,n}_{\psi\psi} - \lambda_{\mp}^{n} }
          { \lambda_{\pm}^{n} - \lambda_{\mp}^{n} }, \nn\\
\Delta \mathcal{B}_{NS}^{n} 
 &=& 24 \frac{n^2 + n + 2}{n(n+1)(n+2)}
        e_{H}^2 ( e_{H}^2 - \langle e^2 \rangle_{n_f} )
	\frac{1}{1 + d_{NS}^{n}}
     \Delta B_{\psi}^{n}, \nn\\
\Delta \mathcal{B}_{\pm}^{n} 
 &=&  24 \frac{n^2 + n + 2}{n(n+1)(n+2)}
        e_{H}^2 \langle e^2 \rangle_{n_f} 
	\frac{1}{1 + d_{\pm}^{n}} \nn\\
&{}& \hspace{0.2cm} 
 \times \left[  
	       \frac{\gamma^{0,n}_{\psi\psi} - \lambda_{\mp}^{n} }
	            { \lambda_{\pm}^{n} - \lambda_{\mp}^{n} }
	       \Delta B_{\psi}^{n}
	     + 
	       \frac{ \gamma^{0,n}_{G\psi} }
	            { \lambda_{\pm}^{n} - \lambda_{\mp}^{n} }
	       \Delta B_{G}^{n}
	\right], \nn\\
\Delta \mathcal{C}^{n}
 &=& \frac{1}{n_f} 12 \beta_0 e_{H}^2 
     \left(  \Delta B_{G}^{n} + \Delta \tilde{A}_{nG}^{\psi}  \right).
\label{master-2}
\end{eqnarray}
where $e_{H}$ is the heavy quark charge,
\begin{eqnarray}
 e_{H} = 
\begin{cases} 
   +\frac{2}{3},  & \mbox{ for $SU(2)_{L}$ up-type quark,} \\
  -\frac{1}{3},  & \mbox{ for $SU(2)_{L}$ down-type quark}~. 
\end{cases}
\end{eqnarray}
Here we have adopted the notation of Ref.~\citen{real_photon_NLO} for
the anomalous dimensions and coefficient functions in the 
$\overline{\rm MS}$ scheme\cite{BBDM}, and $\Delta\tilde{\Av}_n^{(1)}
=6(\langle e^2 \rangle,0,\langle e^4 \rangle-\langle e^2 \rangle^2)
\Delta\tilde{A}_{nG}^{\psi}$.

Thus there is no change for the moment at LO level.
This is explained by the following discussion.
In terms of our approach, the heavy quark effect is included in 
the RGE inputs; OME and the coefficient functions. 
The evolution factor is the same as the massless case and there is 
no physical difference which distinguishes the quarks except for 
the electric charge in the RGE inputs at this order (LO).
This results is also justified by the explicit calculation for 
the LO moment. So, the heavy quark effect is occurred at NLO level.
Therefore the higher order(more than NLO) calculation is essential 
in the case of including the heavy quark effects with massless 
calculations for the moments. The variation of coefficients 
$\Delta \mathcal{A}_{i}^{n}, \Delta \mathcal{B}_{i}^{n}, 
\Delta \mathcal{C}^{n}$ can be obtained by considering the possible
variation of the coefficient functions and the operator matrix element
at NLO. 

Here we confine ourselves to the case of the limit:
$\Lambda_{\rm QCD}^2 \ll P^2 \ll m^2 \ll Q^2$. 
In this limit we have
\begin{eqnarray}
 \Delta \tilde{A}_{nG}^{\psi} \frac{1}{n_f}
 &=& 2\left[ - \frac{n^2+n+2}{n(n+1)(n+2)} 
               \ln \frac{m^2}{P^2}
	     + \frac{1}{n} - \frac{1}{n^2}
\right. \nn \\
&{}& \left.  \hspace{0.5cm}
	     + \frac{4}{(n+1)^2} - \frac{4}{(n+2)^2}
	     - \frac{n^2+n+2}{n(n+1)(n+2)} \sum_{j=1}^n\frac{1}{j}
       \right]~,\\
\Delta B^n_\psi&=&0,\quad \Delta B^n_G=0,\quad 
\Delta B^n_\gamma=2\Delta B^n_G/n_f=0. 
\end{eqnarray}
The quark (gluon) coefficient functions $B^n_\psi$ ($B^n_G$) are
obtained by taking the difference between photon-parton amplitude
and the operator matrix elements\cite{BBDM,Floratos-etal}. 
In the large mass limit 
$P^2 \ll m^2$, the deviations from the massless case are the
same for the photon-parton amplitudes and the operator matrix 
elements. Hence we have $\Delta B^n_\psi=\Delta B^n_G=0$. 
We will discuss the details elsewhere\cite{heavy_quark_effects_full}.
We also note here for the longitudinal structure function
$F_L^\gamma(x,Q^2,P^2)$ we do not have heavy quark mass effects
to the LO (${\cal O}(\alpha)$) and is given by the same 
formula as the massless case.  

\section{Numerical analysis of $F_{\rm eff}^{\gamma}(x,Q^2,P^2)$}
The virtual photon structure functions are recovered from the moments 
by the inverse Mellin transformation.
In this section we examine the heavy quark mass effects on 
the effective photon structure function \cite{F_eff} 
$F_{\rm eff}^{\gamma}(x, Q^2, P^2)$ defined as
\begin{eqnarray}
 F_{\rm eff}^{\gamma}(x, Q^2, P^2) 
  &=&   F_{2}^{\gamma}(x, Q^2, P^2) 
      + \frac{3}{2} F_{L}^{\gamma}(x, Q^2, P^2)~.
\end{eqnarray}
We evaluate $ F_{\rm eff}^{\gamma}$ up to the NLO
and compare our  theoretical predictions with the existing experimental data  
from
PLUTO Collaboration \cite{PLUTO} and L3 Collaboration\cite{L3}. 
For the PLUTO (L3) data, we have $Q^2=5\ (120)\ {\rm GeV}^2$ and $P^2=0.35 
\ (3.7)\ {\rm GeV}^2$. Therefore, we assume that the active flavors are
$u,d,s$ (massless) plus $c$ (heavy) for the case of PLUTO 
and $u,d,s,c$ (massless) plus $b$ (heavy) for L3.

\begin{figure}
 \centerline{\includegraphics[width=12cm,height=8cm]
{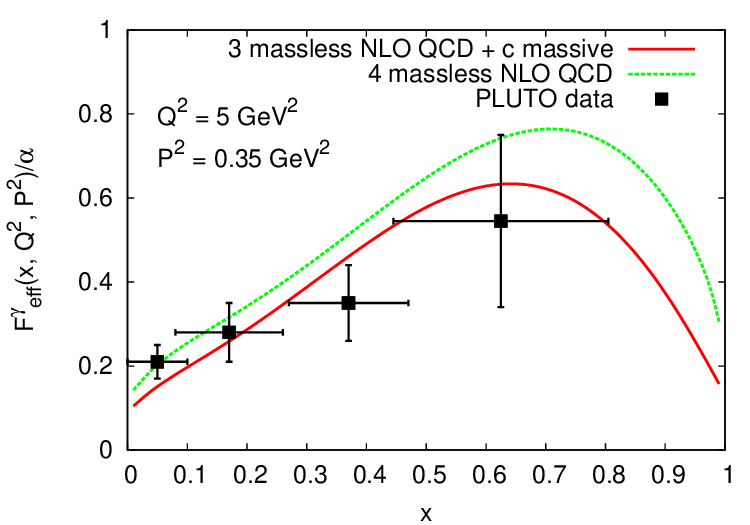}}
\caption{ $F_{\rm eff}^\gamma$ to NLO in QCD and  PLUTO data. 
$n_f=4$, $Q^2=5$GeV$^2$, $P^2=0.35$GeV$^2$, $x_{\max}=0.43$.}
\label{fig2}
\end{figure}

\begin{figure}
 \centerline{\includegraphics[width=12cm,height=8cm]
{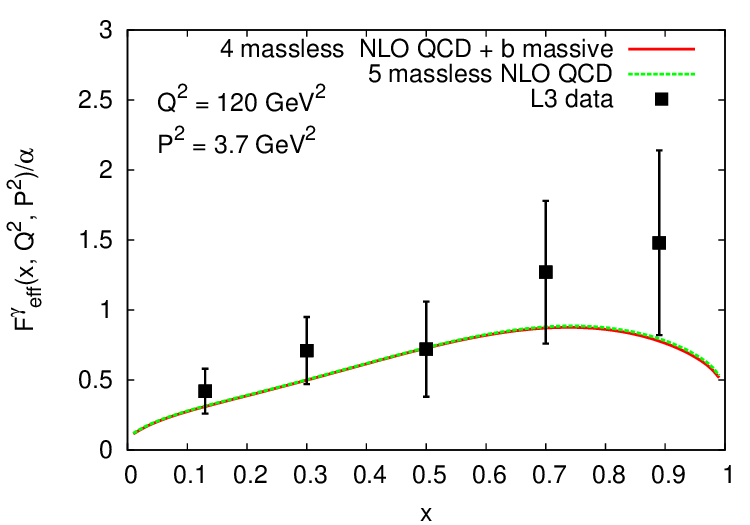}}
\caption{ $F_{\rm eff}^\gamma$ to NLO in QCD and  L3 data. 
$n_f=5$, $Q^2=120$GeV$^2$, $P^2=3.7$GeV$^2$,$x_{\max}=0.63$.}
 \label{fig3}
\end{figure}

We  plot the experimental data from PLUTO group in  Fig.~\ref{fig2} and those 
from L3 group in Fig \ref{fig3}, together with our theoretical predictions. 
We also show the curves of the NLO predictions when active quarks are all 
massless. 
We use the QCD running quark mass ${\overline m}(P^2)$ which is valid up to 
the NLO\cite{gamma_m} and we adopt the following values of the quark masses as 
inputs\cite{PDG}, 
\begin{eqnarray}
 m_{c} &=& 1.3 \mbox{GeV} \hspace{0.5cm}(\mbox{for PLUTO}),\\
 m_{b} &=& 4.2 \mbox{GeV} \hspace{0.5cm}(\mbox{for L3}).
\end{eqnarray}
In general the heavy quark mass  has an effect of reducing the photon
structure functions in magnitude. This feature is explained by the 
suppression of the heavy quark production rate due to the existence of 
their masses. Heavy quark mass effects appear at larger $x$ region. 
Due to the kinematical constraint for the heavy quark production 
$(p+q)^2 \ge 4m^2$, the contribution of heavy quark to the structure 
functions exists below $x_{\max}=\frac{1}{1+\frac{4m^2}{Q^2}}$ and, 
therefore, the difference between the massless and the massive cases 
emerges above $x_{\max}$. This kinematical ``threshold" effect 
is not clearly seen in our analysis since we adopted the framework based 
on the OPE and took into account only the leading twist-2 operators. But 
still we see that the difference between the massless and the massive cases 
becomes bigger at large  $x$ (see Fig.\ref{fig2} and Fig.\ref{fig4} below). 
It is also noted that the heavy quark mass effects are sensitive to the 
electric charge of the relevant quark. 
Since the photon structure functions depend on the quark-charge factors 
$\langle e^2 \rangle$ and $\langle e^4 \rangle$,  the up-type heavy quark
gives larger contribution to the photon structure functions than the down-type quark. 

In the case of PLUTO data,  there is no justification of our approximation
for the limit $Q^2 \gg m^2$.  But we find in Fig.~\ref{fig2} that 
the predicted curve with mass effects shows the trend of reducing
the \lq\lq over-estimated\rq\rq massless QCD calculation, 
especially at larger $x$ region,  and appears to be closer to the experimental 
data. For L3 data, the hierarchical condition $P^2 \ll m^2 \ll Q^2$ is 
satisfied. Although the experimental error bars are rather large, we find 
in Fig.~\ref{fig3} that theoretical curves, both massive and massless cases, 
are  roughly consistent with the data, except for the larger $x$ region. 
For the L3 region, the heavy quark mass effects are almost negligible
since the $b$ has a charge $-1/3$.

Finally we present,  in Fig.~\ref{fig4}, 
our prediction for the case $Q^2=30{\rm GeV}^2$ 
with $P^2=0.35{\rm GeV}^2$, as an illustration when massive charm quark 
is relevant.  The condition $P^2\ll m^2 \ll Q^2$ is satisfied.  
Although there is no  experimental data corresponding to this case, 
we find that the heavy quark mass effects are sizable as can be seen in the 
figure.
\begin{figure}
\centerline{\includegraphics[width=12cm,height=8cm]
{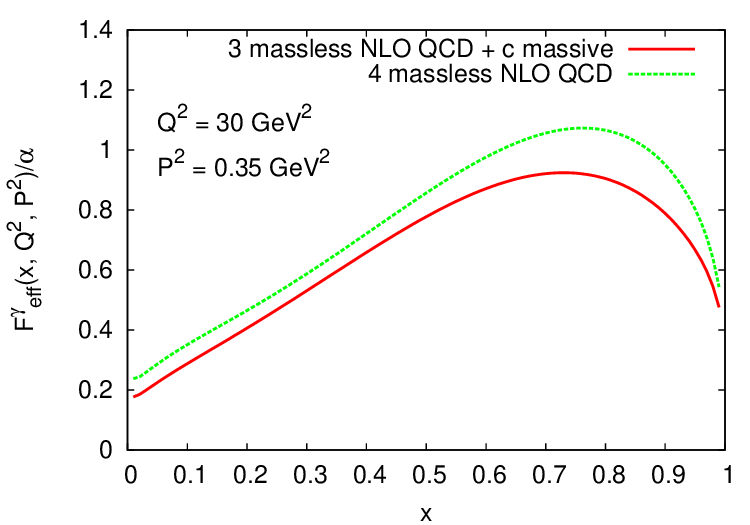}}
\caption{$F_{\rm eff}^{\gamma}$ to NLO in QCD for $n_f=4$ 
$Q^2=30{\rm GeV}^2$ with $P^2=0.35{\rm GeV}^2$, $x_{\max}=0.82$ .}
 \label{fig4}
\end{figure}
\section{Conclusions}
We have investigated the heavy quark mass effects in the
virtual photon structure function based on the parton picture 
as well as on the operator product expansion. 
We have derived the master formula for the additional contributions
at NLO to the moments due to mass effects. 
The heavy quark mass effect does not change the LO moments of
the photon structure functions but it changes NLO moments.
We applied this formalism to the phenomenological analysis of 
$F_{\rm eff}^{\gamma}(x,Q^2,P^2)$. We confronted the theoretical 
QCD prediction to NLO including the heavy quark mass effects with 
the existing experimental data.

For the kinematical region of the PLUTO data we assumed the 3 quarks
(u,d,s) are massless, while the charm quark (c) is the heavy quark.
In the region for the L3 data we took the 4 quarks (u,d,s,c) to
be massless, and the bottom (b) quark to be the heavy quark.
For the PLUTO region, there exists a sizable heavy quark effect at
larger $x$ regime. Although our approximation assuming $m^2 \ll Q^2$
is not immediately applicable for the PLUTO region, the theoretical
prediction shows a right trend of describing the experimental data.   
For the L3 region, the heavy quark mass effects are almost negligible
since the $b$ has a charge $-1/3$.
It is somewhat remarkable that the theoretical predictions for
the both cases are consitent with experimental data as for the
total normalization. We only have one adjustable parameter, 
$\Lambda_{\rm QCD}$, which we took 0.2 GeV.

It would be interesting to investigate the charm and bottom quark mass 
effects at the future SUPER-B experiments \cite{SUPERB}.
If the center of mass energy of the future linear collider (ILC) is 
enough to produce the top quarks, its mass effects 
would be important for the measurements of the virtual photon structure 
functions in view of the charge-factor enhancement. 

\section*{Acknowledgments}
We would like to thank Kiyoshi Kato and Yoshimasa Kurihara for
discussions. This work is supported in part by Grant-in-Aid for
Scientific Research (C) from the Japan Society for the Promotion 
of Science No.18540267.

\appendix
%
\section{Diagonalization of the anomalous dimension matrix}\label{derive1}
We briefly describe the diagonalization of the anomalous dimension matrix
(\ref{anomalous-dim}).
Expanding the anomalous dimension in a power series of the coupling constant 
$g$:
\begin{eqnarray}
\hat{\gamma}(g)=\frac{g^2}{16\pi^2}\hat{\gamma}_n^{(0)}
+\frac{g^4}{(16\pi^2)^2}\hat{\gamma}_n^{(1)}+\cdots~,
\end{eqnarray}
and decomposing the one-loop anomalous dimension matrix $\hat{\gamma}_n^{(0)}$
as a block matrix
\begin{eqnarray}
\hat{\gamma}_n^{(0)}=
\left(
\begin{array}{c|c}
\widetilde{\gamma}_n^{(0)}&0\\
\hline
0&\gamma_{NS}^{0,n}
\end{array}
\right)~,\label{gamma-hat}
\end{eqnarray}
with the $\widetilde{\gamma}_n^{(0)}$ expressed in terms of 
one-loop anomalous dimensions\cite{BBDM}
\begin{eqnarray}
\widetilde{\gamma}_n^{(0)}=
\left(
\begin{array}{ccc}
\gamma_{\psi\psi}^{0,n}\ &0\ &\gamma_{G\psi}^{0,n}\\
0\ &\gamma_{\psi\psi}^{0,n}\ &\gamma_{G\psi}^{0,n}\\
\frac{n_f-1}{n_f}\gamma_{\psi G}^{0,n}\ &
\frac{1}{n_f}\gamma_{\psi G}^{0,n}\ &\gamma_{GG}^{0,n}
\end{array}
\right)~,
\end{eqnarray}
we get the eigenvalues of the above 3 $\times$ 3 matrix, 
$\lambda^n=\lambda_\psi^n,\lambda_{+}^n,\lambda_{-}^n$
given by
\begin{eqnarray}
\lambda_\psi^n=\gamma_{\psi\psi}^{0,n},\ 
\lambda_\pm^n=\frac{1}{2}\left\{
\gamma_{\psi\psi}^{0,n}+\gamma_{GG}^{0,n}\pm
\sqrt{(\gamma_{\psi\psi}^{0,n}-\gamma_{GG}^{0,n})^2+4\gamma_{\psi G}^{0,n}
\gamma_{G\psi}^{0,n}}\right\}~.
\end{eqnarray}

Introducing the projection operators we can write down 
$\widetilde{\gamma}_n^{(0)}$ as
\begin{eqnarray}
\widetilde{\gamma}_n^{(0)}=\sum_{i=\psi,\pm}\lambda_i^n P_i^n~,
\end{eqnarray}
where the projection operators are given as
\begin{eqnarray}
P_\psi^n&=&
\left(
\begin{array}{ccc}
\frac{1}{n_f}\ &-\frac{1}{n_f}\ &0\\
-\frac{n_f-1}{n_f}\ &\frac{n_f-1}{n_f}\ &0\\
0&0&0
\end{array}
\right)~,\\
P_\pm^n&=&
\frac{1}{\lambda_\pm^n-\lambda_\mp^n}
\left(
\begin{array}{ccc}
\frac{n_f-1}{n_f}(\gamma_{\psi\psi}^{0,n}-\lambda_\mp^n) &
\frac{1}{n_f}(\gamma_{\psi\psi}^{0,n}-\lambda_\mp^n)\ &\gamma_{G\psi}^{0,n}\\
\frac{n_f-1}{n_f}(\gamma_{\psi\psi}^{0,n}-\lambda_\mp^n) &
\frac{1}{n_f}(\gamma_{\psi\psi}^{0,n}-\lambda_\mp^n)\ &\gamma_{G\psi}^{0,n}\\
\frac{n_f-1}{n_f}\gamma_{\psi G}^{0,n}\ &\frac{1}{n_f}\gamma_{\psi G}^{0,n}
&\gamma_{GG}^{0,n}-\lambda_\mp^n
\end{array}
\right)~.
\end{eqnarray}
Including flavor 
non-singlet anomalous dimension $\gamma_{NS}^n$ 
into the $4\times 4$ matrix (\ref{gamma-hat}),
the projection operators are extended to
\begin{eqnarray}
\hat{\gamma}_n^{(0)}=\sum_{i=\psi,+,-,NS}\lambda_i^n P_i^n~,
\end{eqnarray}
where
\begin{eqnarray}
&&\hspace{-0.5cm}P_\psi^n=
\left(
\begin{array}{ccc|c}
\frac{1}{n_f}\ &-\frac{1}{n_f}\ &0&0\\
-\frac{n_f-1}{n_f}\ &\frac{n_f-1}{n_f}\ &0&0\\
0&0&0&0\\
\hline
0&0&0&0
\end{array}
\right)~,\\
&&\hspace{-0.5cm}P_\pm^n=
\frac{1}{\lambda_\pm^n-\lambda_\mp^n}
\left(
\begin{array}{ccc|c}
\frac{n_f-1}{n_f}(\gamma_{\psi\psi}^{0,n}-\lambda_\mp^n) &
\frac{1}{n_f}(\gamma_{\psi\psi}^{0,n}-\lambda_\mp^n)\ &\gamma_{G\psi}^{0,n}&0\\
\frac{n_f-1}{n_f}(\gamma_{\psi\psi}^{0,n}-\lambda_\mp^n) &
\frac{1}{n_f}(\gamma_{\psi\psi}^{0,n}-\lambda_\mp^n)\ &\gamma_{G\psi}^{0,n}&0\\
\frac{n_f-1}{n_f}\gamma_{\psi G}^{0,n}\ &\frac{1}{n_f}\gamma_{\psi G}^{0,n}
&\gamma_{GG}^{0,n}-\lambda_\mp^n&0\\
\hline
0&0&0&0
\end{array}
\right)~,\\
&&\hspace{-0.5cm}P_{NS}^n=
\left(
\begin{array}{ccc|c}
0\  &0\ &0\ &0 \\
0\  &0\ &0\ &0 \\
0\  &0\ &0\ &0 \\
\hline
0&0&0&1
\end{array}
\right)~.
\end{eqnarray}
Now let us see that the leading order coefficients ${\cal L}_i$
are not changed in the presence of the heavy quark mass effects.
With the four eigenvalues, the leading-order coefficients $\widehat{\cal L}_i$
turn out to be
\begin{eqnarray}
&&\widehat{\cal L}_\psi^n=24\frac{n^2+n+2}{n(n+1)(n+2)}
\frac{n_f-1}{n_f}(\langle e^2\rangle_{n_f-1}-e_H^2)^2
\frac{1}{1+d_\psi^n}~,\\
&&\widehat{\cal L}_\pm^n=24\frac{n^2+n+2}{n(n+1)(n+2)}
\frac{1}{\lambda_\pm^n-\lambda_\mp^n}\frac{1}{n_f}
\left\{(n_f-1)\langle e^2\rangle_{n_f-1}+e_H^2\right\}^2\nonumber\\
&&\hspace{2cm}\times(\gamma_{\psi\psi}^{(0,n)}-\lambda_\mp^n)
\frac{1}{1+d_\pm^n}~,\nonumber\\
&&\widehat{\cal L}_{NS}^n=24\frac{n^2+n+2}{n(n+1)(n+2)}(n_f-1)
(\langle e^2\rangle_{n_f-1}-e_H^2)^2\frac{1}{1+d_{NS}^n}~,
\end{eqnarray}
where we have denoted the LO coefficients with a hat for the 
massive case in order to distinguish them from those for the massless case.
Hence the leading-order QCD result for the moment of $F_2^\gamma$
is given by
\begin{eqnarray}
&&\int_0^1 dx \, x^{n-2} F_2^\gamma(x,Q^2,P^2) \nonumber\\
&&=\frac{\alpha}{4\pi} \frac{1}{2\beta_0}
\sum_{i=\psi,\pm,NS} \widehat{\cal L}^n_i \frac{4\pi}{\alpha_s(Q^2)} 
\left[ 1 - \left(\frac{\alpha_s(Q^2)}{\alpha_s(P^2)}\right)^{d^n_i+1}\right] ~.
\nonumber\\
\end{eqnarray} 
Here we note that
\begin{eqnarray}
\lambda_\psi=\lambda_{NS}=\gamma_{\psi\psi}^{0,n},\quad
d_\psi^n=d_{NS}^n=\gamma_{\psi\psi}^{0,n}/2\beta_0~.
\end{eqnarray}
So we find the sum of $\widehat{\cal L}_\psi^n$ and $\widehat{\cal L}_{NS}^n$:
\begin{eqnarray}
&&\widehat{\cal L}_\psi^n+\widehat{\cal L}_{NS}^n=
24\frac{n^2+n+2}{n(n+1)(n+2)}\frac{1}{1+d_\psi^n}\nonumber\\
&&\times\left\{\frac{n_f-1}{n_f}(\langle e^2\rangle_{n_f-1}-e_H^2)^2~,
+(n_f-1)\langle e^4\rangle_{n_f-1}-(n_f-1)
\langle e^2\rangle_{n_f-1}^2\right\}~.\end{eqnarray}
Now let us remind the following relations:
\begin{eqnarray}
&&\langle e^2\rangle_{n_f-1}
\equiv \frac{1}{n_f-1}\sum_{i=1}^{n_f-1}e_i^2,\quad 
\langle e^4\rangle_{n_f-1}
\equiv \frac{1}{n_f-1}\sum_{i=1}^{n_f-1}e_i^4,
\end{eqnarray}
then we have
\begin{eqnarray}
&&\widehat{\cal L}_\psi^n+\widehat{\cal L}_{NS}^n=
24\frac{n^2+n+2}{n(n+1)(n+2)}\frac{1}{1+d_\psi^n}\nonumber\\
&&\times\left\{
\sum_{i=1}^{n_f}e_i^4-\frac{1}{n_f}\left(
\sum_{i=1}^{n_f}e_i^2-e_H^2\right)^2-\frac{2}{n_f}e_H^2
\left(\sum_{i=1}^{n_f}e_i^2-e_H^2\right)-\frac{1}{n_f}e_H^4
\right\}\nonumber\\
&&=24\frac{n^2+n+2}{n(n+1)(n+2)}\frac{1}{1+d_\psi^n}
\left\{\sum_{i=1}^{n_f}e_i^4-\frac{1}{n_f}\left(
\sum_{i=1}^{n_f}e_i^2\right)^2\right\}={\cal L}_{NS}~.
\end{eqnarray}
We also find $\widehat{\cal L}_\pm={\cal L}_\pm$. 
So to the leading-order we get
\begin{eqnarray}
&&\int_0^1 dx \, x^{n-2} F_2^\gamma(x,Q^2,P^2)~, \nonumber\\
&&=\frac{\alpha}{4\pi} \frac{1}{2\beta_0}
\sum_{i=\pm,NS} {\cal L}^n_i \frac{4\pi}{\alpha_s(Q^2)} 
\left[ 1 - \left(\frac{\alpha_s(Q^2)}{\alpha_s(P^2)}\right)^{d^n_i+1}\right]~,
\nonumber\\
\end{eqnarray}
which is nothing but the expression for the $n_f$ light-flavor
case as we expected. Note that in the above equation, we have
used the fact:
\begin{eqnarray}
\sum_{i=1}^{n_f-1}e_i^2+e_H^2=\sum_{i=1}^{n_f}e_i^2,\quad
\sum_{i=1}^{n_f-1}e_i^4+e_H^4=\sum_{i=1}^{n_f}e_i^4~.
\end{eqnarray}
The above result means that there is no difference between the case with
$n_f-1$ light-flavor plus one heavy-flavor and the one with
$n_f$ light-flavors to the leading-order in QCD. For the NLO coefficients
$\Delta \mathcal{A}_{i}^{n}, 
\Delta \mathcal{B}_{i}^{n}, \Delta \mathcal{C}^{n}$
we perform the similar analysis and get the additional contributions
given in Eq.~(\ref{master-2}).

\section{Master formula for the moments in the parton picture}\label{derive2}
There is an alternative method to derive the moment sum rule Eq.(\ref{master})
 which is based on the parton picture \cite{PDF_decomposition}.
Consider the $n$-th moment of the virtual photon structure
function $F_{2}^{\gamma}(x, Q^2, P^2)$ in the case where $n_f-1$ light
quarks and one heavy quark are present. The $n$-th moments of $F_{2}^{\gamma}$ 
is given by
\begin{eqnarray}
 M_{2}^{\gamma}(n) 
  =   q_{L}^{\gamma}(n)  C^{L}(n)
         + q_{H}^{\gamma}(n)  C^{H}(n) 
         + G^{\gamma}(n)      C^{G}(n)
	 + q_{NS}^{\gamma}(n) C^{NS}(n)
	 + C^{\gamma}(n)~, \nn\\
\end{eqnarray}
where we have suppressed, for simplicity, 
the $Q^2$ as well as $P^2$ dependence of the moments of the structure function,
the parton distributions and the coefficient functions. 
$q_{L(NS)}^{\gamma}$ denotes the flavor singlet (non-singlet) quark parton 
distribution function for
the $n_f - 1$ flavors as defined in Eq.(\ref{singlet-NS}), 
and $G^{\gamma}$ is the gluon parton distribution function.
$C^{i}(i=L,H,G,NS,\gamma)$ are the coefficient functions for
parton $i$-type in the virtual photon,
\begin{eqnarray}
 &&C^{L}(n) = \langle e^2 \rangle_{n_f-1}
              \left( 1 + \frac{\alpha_s}{4\pi} B_{\psi}^{n} \right),\\
 &&C^{NS}(n) = 1 + \frac{\alpha_s}{4\pi} B_{\psi}^{n},\\
 &&C^{H}(n) = C^{L} + e_{H}^2 \frac{\alpha_s}{4\pi} \Delta B_{\psi}^{n},\\
 &&C^{G}(n) = \langle e^4 \rangle_{n_f-1}
              \frac{\alpha_s}{4\pi} B_{G}^{n},\\
&& C^{\gamma}(n) = 2\beta_{0} 
               \left\{   \delta_{\gamma} B_{\gamma}^{n} 
		      + 3e_{H}^4( B_{\gamma}^{n} + \Delta B_{\gamma}^{n} )
	       \right\}.
\end{eqnarray}
Putting all the above quantities together and noting that the heavy quark 
distribution differs from the light-flavor distribution by 
an extra contribution $\Delta  q^{n_f}(n)$:
\begin{eqnarray}
 q^{H}(n) &=& q^{n_f}(n) + \Delta  q^{n_f}(n)~,
\end{eqnarray}
we obtain the moment which includes the heavy quark mass effects as
\begin{eqnarray}
 M^{\gamma}(n) 
  &=&   M^{\gamma}(n) \mid_{m=0} 
      + e_{H}^2 \Delta q^{n_f}(n)
      + 6 \beta_0 e_{H}^4 \Delta B_{\gamma}^{n}
\nn\\
&{}& \hspace{0.2cm}
      + e_{H}^2 \frac{\alpha_s}{4\pi} q^{n_f}(n) \Delta B_{\gamma}^{n}
      + e_{H}^2 \frac{\alpha_s}{4\pi} G^{\gamma}(n) 
        \frac{1}{n_f} \Delta B_{G}^{n}~.
\end{eqnarray}
Here we note that $\Delta B^n_\gamma=2\Delta B_G^n/n_f$. Denoting 
$r=\alpha_s(Q^2)/\alpha_s(P^2)$ we have
\begin{eqnarray}
&&\Delta q^{n_f}(n)=\frac{e_H^2-\langle e^2\rangle}
{n_f(\langle e^4\rangle -\langle e^2\rangle^2)}
\Delta q_{NS}^\gamma(n)+\frac{1}{n_f}\Delta q_S^\gamma(n)~,\\
&&\Delta q_{NS}^\gamma(n)/\frac{\alpha}{8\pi\beta_0}
=\Delta{\cal A}_{NS}^n(1-r^{d^n_{NS}})+\Delta\widetilde{C}_{NS}^n~,\\
&&\Delta q_{S}^\gamma(n)/\frac{\alpha}{8\pi\beta_0}
=\Delta\hat{\cal A}_{S}^{+n}(1-r^{d^n_{+}})+
\Delta\hat{\cal A}_{S}^{-n}(1-r^{d^n_{-}})
+\Delta\hat{C}_{S}^n~,
\end{eqnarray}
where 
\begin{eqnarray}
&&\Delta{\cal A}_{NS}^n=-2\beta_0\Delta A_n^{(2)NS},\quad
\Delta\hat{\cal A}_{S}^{\pm n}=-2\beta_0\frac{\gamma_{\psi\psi}^{0,n}
-\lambda_{\mp}^n}{\lambda_{\pm}^n-\lambda_{\mp}^n}\Delta A_n^{(2)\psi}~,
\nonumber\\
&&\Delta\widetilde{C}_{NS}^n=2\beta_0\Delta A_n^{(2)NS}, \quad
\Delta\hat{C}_{S}^n=2\beta_0\Delta A_n^{(2)\psi}~,
\end{eqnarray}
with
\begin{eqnarray}
&&2\beta_0 A_n^{(2)NS}=12\beta_0\widetilde{A}_{nG}^\psi(
\langle e^4\rangle-\langle e^2\rangle^2)~,\nonumber\\
&&2\beta_0A_n^{(2)\psi}=12\beta_0\widetilde{A}_{nG}^\psi\langle e^2\rangle~.
\end{eqnarray}
The gluon distribution is given by
\begin{eqnarray}
G^\gamma(n)/\frac{\alpha}{8\pi\beta_0}
=\frac{4\pi}{\alpha_s}{\cal L}^{+n}_G(1-r^{d^n_{+}+1})
+\frac{4\pi}{\alpha_s}{\cal L}^{-n}_G(1-r^{d^n_{-}+1})~,
\end{eqnarray}
where
\begin{eqnarray}
{\cal L}^{\pm n}_G=\frac{K^{0,n}_\psi\gamma_{G\psi}^{0,n}}
{\lambda^n_\pm-\lambda^n_\mp}\frac{1}{1+d^n_\pm},\quad
K^{0,n}_\psi=24n_f\langle e^2\rangle_{n_f}\frac{n^2+n+2}{n(n+1)(n+2)}~.
\end{eqnarray}
From these expressions we can derive the (\ref{master}) with (\ref{master-2}).

\newpage

\end{document}